\documentclass[aps,preprint,showpacs,superscriptaddress]{revtex4-1}  % for double-spaced preprint
\usepackage{graphicx}
\usepackage{bm}   
\usepackage{color}
\usepackage{amssymb}
\usepackage{amsmath}
\usepackage{balance}
\usepackage{subfigure}
\newcommand{\mathsym}[1]{{}}
\newcommand{\unicode}[1]{{}}
\usepackage{setspace}

\begin{document}
	
	\title{Nth order smooth positon and breather-positon solutions of a generalized nonlinear Schr\"{o}dinger equation}
	\author{N. Vishnu Priya}
	\affiliation{Department of Mathematics, Indian Institute of Science, Bangalore-560012, Karnataka, India.}
	\author{S. Monisha}
	\affiliation{Department of Nonlinear Dynamics, Bharathidasan University, Tiruchirappalli - 620 024, Tamilnadu, India.}
	\author{M. Senthilvelan}
	\email[Correspondence to: ]{velan@cnld.bdu.ac.in}
	\affiliation{Department of Nonlinear Dynamics, Bharathidasan University, Tiruchirappalli - 620 024, Tamilnadu, India.}
	\author{Govindan Rangarajan}
	\affiliation{Department of Mathematics, Indian Institute of Science, Bangalore-560012, Karnataka, India.}
	
	\begin{abstract}
	In this paper, we investigate smooth positon and breather-positon solutions of a generalized nonlinear Schr\"{o}dinger (GNLS) equation which contains higher order nonlinear effects.  With the help of generalized Darboux transformation (GDT) method we construct $N$th order smooth positon solutions of GNLS equation.  We study the effect of higher order nonlinear terms on these solutions.  Our investigations show that the positon solutions are highly compressed by higher order nonlinear effects.  The direction of positons are also get changed.  We also derive $N$th order breather-positon (B-P) solution with the help of GDT.  We show that these B-Ps are well compressed by the effect of higher order nonlinear terms but the period of B-P solution is not affected as in the breather solution case.
	\end{abstract}
	
	%\maketitle

	\maketitle
\section{Introduction}
Studies on nonlinear evolution equations are of contemporary interest in several areas of physics.  One of the most familiar models among nonlinear evolution equations is the Kortweg-deVries (KdV) equation.   This equation appears in shallow water surfaces, internal waves, ion-acoustic waves in plasmas, acoustic waves in a harmonic crystal and so on \cite{Ablowitz}.  This equation has been studied  thoroughly in the literature and has been shown to possess several novel  multisoliton solutions, including positons, negatons and rational solutions \cite{{Drazin},{Matveev1},{Matveev2},{zowr},{Matveev3},{Matveev4}}, to name a few.  The positon type solutions for the KdV equation, using generalized Darboux transformation (DT) method, was first given by Matveev \cite{Matveev1}.  These positons have oscillating nature and appear as slowly decreasing singular solutions which are long range analogues of  solitons \cite{Matveev3}. . In particular, soliton solutions of KdV equation can be obtained by solving its Lax pair equations with negative eigenvalues. The soliton solutions of KdV equation can be obtained by solving its Lax pair equations with negative eigenvalues.  In other words, solitons are associated with negative energy bound states of the Schr\"{o}dinger operator.  On the other hand if we choose positive sign for the eigenvalue parameter then we end up with periodic solutions which are not of physical interest.  However, by expanding these periodic solutions at a particular eigenvalue with Taylor series and substituting them in the Wronskian representations of solutions of KdV equation we can obtain positon solutions.  Hence these positons are generated by positive spectral singularity embedded in the continuous spectrum \cite{Matveev2}.  Subsequently a detailed study on the dynamics of positon solutions of KdV equation has been reported in \cite{Maich}. 

\par These positon solutions may serve as a prototype for shallow water rogue waves (RWs) \cite{Matveev4}.  For the past two decades physicists have carried out intensive research on RWs \cite{{Onorato},{Kharif}}.  These RWs are extreme wave events which make major destructive effects in ocean.  However it has potential applications in nonlinear optical fibers \cite{{nlo1},{nlo2}}, Bose-Einstein condensates \cite{bec}, super fluid He \cite{He} and plasma physics \cite{plasma}.  By considering the importance of positon solutions, efforts have been made to identify singular positon solutions in many nonlinear integrable evolution equations including modified KdV equation \cite{mkdv}, Toda chain \cite{Toda}, sine-Gordon equation \cite{sineg}, KdV and modified KdV hierarchies \cite{Avinash}, fifth order KdV equation \cite{fKdV}, extended KdV equation \cite{eKdV} and KdV equation with self-consistent sources \cite{sKdV}.   Non-singular positon solutions on vanishing background are named as smooth positons or degenerate soliton solutions \cite{{jing},{bk},{ckdv},{comKdv},{dnls},{kundu},{cmkdv}}.  This kind of solution has also been constructed for several nonlinear partial differential equations including nonlinear Schr\"{o}dinger (NLS) equation \cite{{Akhmediev1},{jing}},  Bogoyavlensky–Konoplechenko equation \cite{bk}, coupled KdV \cite{ckdv} and mKdV equations \cite{comKdv}, derivative NLS equation \cite{dnls}, nonlocal Kundu-NLS equation \cite{kundu}, complex mKdV equation \cite{cmkdv}, Wadati-Konno-Ichikawa equation \cite{Wadati} and higher-order Chen-Lee-Liu equation \cite{cll}.  Recently positons on nonvanishing background for the NLS equation have also been constructed and they have been coined as breather-positon (B-P) solutions.  These solutions are also being designated as degenerate breathers in the literature \cite{{Akhmediev1},{Akhmediev2}}.  The central region of B-P solution exhibits a similar structure as that of rogue wave pattern.  Studies on B-P solutions have been made on complex mKdV equation \cite{cmkdv2}, Kundu-Eckhaus equation \cite{Kundu2}, Sasa-Satsuma equation \cite{sasa}, NLS-Maxwell Bloch equations \cite{nlsmb}.  However, positon solutions with higher order nonlinear effects are less studied. 

\par In this paper, we consider a generalized nonlinear Schr\"{o}dinger equation which is of the form
\begin{eqnarray}
	iq_t+q_{xx}+2|q|^2q+\nu(q_{xxxx}+8|q|^2q_{xx}+2q^2q^*_{xx}+4q|q_x|^2+6q^{2}_xq^*+6|q|^4q)=0,
	\label{eq1}
\end{eqnarray}
where $q$ is the amplitude of slowly varying pulse envelope.  $x$ and $t$ are space and time variables respectively.  $*$ denotes complex conjugate. The choice $\nu=0$ in Eq. (1) provides the standard NLS equation. The complete integrability of NLS equation was first shown by Zakharov and Shabat through inverse scattering transform method \cite{Zakharov}. They have also shown that the NLS equation has an infinite number of integrals of motion and possesses N-soliton solutions and their interaction is elastic. Equation (\ref{eq1}) describes the nonlinear wave propagation in various physical systems.  In nonlinear optics Eq. (\ref{eq1}) models the propagation of ultrashort pulses which are responsible for higher-order dispersion and higher order nonlinear effects such as self-steepening, self-frequency and quintic effects \cite{porsezian1}.   This equation also models the dynamics of Heisenberg ferromagnetic spin chain in continuum limit \cite{porsezian2} and higher order excitation of alpha-helical protein \cite{alpha}.  A family of solutions including multisoliton solutions, higher order breather and rogue wave solutions have been constructed for the Eq. (\ref{eq1}) \cite{{porsezian1},{porsezian2}}.  Second order B-P solution of (\ref{eq1}) can be found in \cite{Akhmediev2}.  However, $N$th order smooth positon solutions and $N$th order B-P solutions and the effect of higher order nonlinear terms on these solutions have not been reported.  It will be an interesting problem to investigate how these higher order nonlinear terms affect the smooth positons and B-P solutions by changing the value of the parameter $\nu$.  The prime aim of this paper is to answer to this question.  Such a study not only helps to understand further about the optical system (\ref{eq1}) but also provides an additional view on the associated biological and spin analogues of (\ref{eq1}). 

\par We organize our work as follows. In Sec.2, to begin, we recall the method of deriving positon solutions of KdV equation.  We present $N$th iterated DT solution formula of GNLS equation and explain the method of obtaining $N$th order solution formula for positon solutions of GNLS equation in Sec.3.  We demonstrate the method of constructing smooth positon solutions of GNLS equation in Sec.4 and study the effect of higher order nonlinear terms on these solutions.  In Sec.5 we derive B-P solutions of GNLS equation and study how the B-Ps are affected by higher order nonlinear terms.  Finally, we give our conclusions in Sec.6.  
\section{Positons of KdV equation}
As we mentioned earlier, positons of KdV equation were constructed in \cite{Matveev1}.  For the purpose of reader's quick grasping, in the following, we briefly recall the essential steps involved in the derivation of positon solutions of KdV equation, that is
\begin{eqnarray}
	u_t-6uu_x+u_{xxx}=0.
	\label{eq2}
\end{eqnarray}  
The Lax pair of KdV equation is given by
\begin{eqnarray}
	-\phi_{xx}+u\phi=\lambda\phi,\label{eq3e}\nonumber\\
	\phi_t=-4\phi_{xxx}+6u\phi_x+3u_x\phi.
	\label{eq3}
\end{eqnarray}
If we fix $\phi_1$, $\phi_2$,...,$\phi_N$ are solutions of Eq.(\ref{eq3}) at $\lambda=\lambda_1,\lambda_2,...,\lambda_N$ respectively, then the Nth iterated DT produces new eigenfunctions in the form \cite{Matveev1}
\begin{eqnarray}
	\phi[N]=\frac{W(\phi_1,\phi_2,...,\phi_N,\phi)}{W(\phi_1,\phi_2,...,\phi_N)},
	\label{eq4}
\end{eqnarray}
with new solutions
\begin{eqnarray}
	u[N]=u-2\partial_X^2\ln W(\phi_1,\phi_2,...,\phi_N).
	\label{eq5}
\end{eqnarray}
\par The generalization of the above Wronskian formula can be rewritten to include the degenerate solutions of KdV equation.  This procedure is also called as generalized DT (GDT) method.  In the GDT, the Nth iterated GDT gives new eigenfunction of the form    
\begin{equation}
	\phi[N]=\frac{W(\phi_1,\partial_\lambda\phi_1,\partial_{\lambda}^2\phi_1,...,\partial_{\lambda}^{m_1}\phi_1,\phi_2,\partial_\lambda\phi_2,\partial_{\lambda}^2\phi_2,...,\partial_{\lambda}^{m_2}\phi_2,...,\phi_N,\partial_{\lambda}\phi_N,\partial_{\lambda}^2\phi_N,...,\partial_{\lambda}^{m_N}\phi_N,\phi)}{W(\phi_1,\partial_\lambda\phi_1,\partial_{\lambda}^2\phi_1,...,\partial_{\lambda}^{m_1}\phi_1,\phi_2,\partial_\lambda\phi_2,\partial_{\lambda}^2\phi_2,...,\partial_{\lambda}^{m_2}\phi_2,...,\phi_N,\partial_{\lambda}\phi_N,\partial_{\lambda}^2\phi_N,...,\partial_{\lambda}^{m_N}\phi_N)}.
	\label{eq6}
\end{equation}
The Nth iterated new solutions are given by \cite{Matveev1}
\begin{eqnarray}
	u[N]=u-2\partial_X^2\ln W(\phi_1,\partial_\lambda\phi_1,...,\partial_{\lambda}^{m_1}\phi_1,\phi_2,\partial_\lambda\phi_2,...,\partial_{\lambda}^{m_2}\phi_2,...,\phi_N,\partial_{\lambda}\phi_N,...,\partial_{\lambda}^{m_N}\phi_N),
	\label{eq7}
\end{eqnarray}
where $m_i,\ i=1,2,...,N,$ can be chosen arbitrarily.  If we assume the seed solution as $u=0$ at $\lambda=\lambda_1$ and solving the Lax pair equations (\ref{eq3e}) we obtain $\phi_1=c_1e^{i\sqrt{\lambda_1}x+4i\lambda_{1}^{3/2}t}+c_2e^{-i\sqrt{\lambda_1}x-4i\lambda_{1}^{3/2}t}$, where $c_1$ and $c_2$ are integration constants.  By choosing a negative constant for the eigenvalue such as $\lambda_1=-k_1^2$ with $c_1=c_2=\frac{1}{2}$, $\phi_1$ becomes a hyperbolic cosine function, that is $\cosh(k_1x-4k_1^3t)$.  Substituting this $\phi_1$ in the solution formula (\ref{eq5}) with $N=1$ we can obtain one soliton solution of KdV equation.  In contrast, if we consider a positive eigenvalue, that is $\lambda_1=k_1^2$ then $\phi_1$ takes trigonometric cosine function, that is $\phi_1=\cos(k_1x+4k_1^3t)$.  Now plugging this function in (\ref{eq4}) with $N=1$ we can obtain the periodic solution of KdV equation which is not of much interest.  But by substituting this cosine function in the solution formula (\ref{eq7}) we can produce $N$th order positon solution of KdV equation.  For different eigenvalues $(\lambda_2,...,\lambda_N)$, Eq. (\ref{eq6}) gives positons of higher generation.  In the next section, we discuss how this procedure can be extended to GNLS equation. 
\section{DT of GNLS Eq.(\ref{eq1})}
Now we focus our attention on Eq. (\ref{eq1}).  The Lax pair of GNLS equation is given by \cite{porsezian1}
\begin{eqnarray}
	\Phi_x=U\Phi,\ \Phi_t=V\Phi.
	\label{eq8}
\end{eqnarray}
where
\begin{eqnarray}
	U&=&Q-i\lambda\sigma_3,\nonumber\\
	V&=&(3i\nu|q|^4+i|q|^2+i\nu(q^*q_{xx}+qq^{*}_{xx}-|q_x|^2)-2\lambda\nu(qq^{*}_x-q_xq^*)-2i\lambda^2(2\nu|q|^2+1)+8i\lambda^4\nu)\sigma_3\nonumber\\
	&&-8\nu\lambda^3Q-4i\nu\lambda^2\sigma_3Q_x+6i\nu Q^2Q_x\sigma_3+i\sigma_3Q_x+i\nu\sigma_3Q_{xxx}+2\lambda(Q+\nu Q_{xx}-2\nu Q^3),
	\label{eq9}
\end{eqnarray}
and
\begin{eqnarray}
	Q=\begin{pmatrix}0&q\\-q^*&0\end{pmatrix},\;\ \sigma_3=\begin{pmatrix}1&0\\0&-1\end{pmatrix}, 
	\label{eq10}
\end{eqnarray}
where $\lambda$ is a spectral parameter and $\Phi(x,t,\lambda)=(\psi,\phi)^T$ is the two component vector.  The zero curvature condition $U_t-V_x+[U,V]=0$, where the square bracket denotes the usual commutator, gives Eq. (\ref{eq1}).   
\par DT is one of the versatile tools to derive a variety of solutions for an integrable equation. To learn the utility of DT method and its applications in connection with nonlinear evolutionary equations one may refer the book by Matveev and Salle \cite{Matveevbook}. DT for the GNLS Eq. (\ref{eq1}) has already been constructed in the literature \cite{porsezian2}.  Hence, in the following, we present only the essential steps which we need for the present work.  If we consider $\Phi_i,\;i=1,2,...,N,$ as solutions of the Lax pair equations (\ref{eq8}) at $\lambda_i,\;i=1,2,...,N,$ then the N-fold DT gives solution of GNLS equation which can be represented in the form
\begin{eqnarray}
	q[N]=q-2i\frac{|D_{1}^{[N]}|}{|D_{2}^{[N]}|},
	\label{eq11}
\end{eqnarray}
where
\begin{eqnarray}
	D_{1}^{[N]}=\begin{pmatrix}\psi_1&\phi_{1}&\lambda_{1}\psi_1&\lambda_{1}\phi_1&\lambda_1^2\psi_1&\lambda_{1}^2\phi_1&\cdots&\lambda_{1}^{N-1}\psi_1&\lambda_1^N\psi_1\\-\phi_1^*&\psi_{1}^*&-\lambda_{1}^{*}\phi_1^*&\lambda_{1}^{*}\psi_1^*&-\lambda_1^{*2}\phi_1^*&\lambda_{1}^{*2}\psi_1^*&\cdots&-\lambda_{1}^{*N-1}\phi_1^*&-\lambda_{1}^{*N}\phi_1^*\\\psi_2&\phi_{2}&\lambda_{2}\psi_2&\lambda_{2}\phi_2&\lambda_2^2\psi_2&\lambda_{2}^2\phi_2&\cdots&\lambda_{2}^{N-1}\psi_2&\lambda_2^N\psi_2\\
		\vdots&\vdots&\vdots&\vdots&\vdots&\vdots&\cdots&\vdots&\vdots\\\psi_N&\phi_{N}&\lambda_{N}\psi_N&\lambda_{N}\phi_N&\lambda_N^2\psi_N&\lambda_{N}^2\phi_N&\cdots&\lambda_{N}^{N-1}\psi_N&\lambda_N^N\psi_N\end{pmatrix}
	\label{eq12}
\end{eqnarray}
and
\begin{eqnarray}
	D_{2}^{[N]}=\begin{pmatrix}\psi_1&\phi_{1}&\lambda_{1}\psi_1&\lambda_{1}\phi_1&\lambda_1^2\psi_1&\lambda_{1}^2\phi_1&\cdots&\lambda_{1}^{N-1}\psi_1&\lambda_1^{N-1}\phi_1\\-\phi_1^*&\psi_{1}^*&-\lambda_{1}^{*}\phi_1^*&\lambda_{1}^{*}\psi_1^*&-\lambda_1^{*2}\phi_1^*&\lambda_{1}^{*2}\psi_1^*&\cdots&-\lambda_{1}^{*N-1}\phi_1^*&\lambda_{1}^{*N-1}\psi_1^*\\\psi_2&\phi_{2}&\lambda_{2}\psi_2&\lambda_{2}\phi_2&\lambda_2^2\psi_2&\lambda_{2}^2\phi_2&\cdots&\lambda_{2}^{N-1}\psi_2&\lambda_2^{N-1}\phi_2\\
		\vdots&\vdots&\vdots&\vdots&\vdots&\vdots&\cdots&\vdots&\vdots\\\psi_N&\phi_{N}&\lambda_{N}\psi_N&\lambda_{N}\phi_N&\lambda_N^2\psi_N&\lambda_{N}^2\phi_N&\cdots&\lambda_{N}^{N-1}\psi_N&\lambda_N^{N-1}\phi_N\end{pmatrix}.
	\label{eq13}
\end{eqnarray}
\par Using the solution formula (\ref{eq11}) one can construct various solutions of GNLS equation (\ref{eq1}) such as solitons, breathers and rogue wave solutions.  For example, by choosing zero seed solution, that is, $q=0$ and solving Lax pair equations (\ref{eq8}) with $N$-eigenvalues one can obtain $N$-eigenfunctions.  Substituting these eigenfunctions in the solution formula (\ref{eq11}) one can get $N$-soliton solutions of GNLS equation (\ref{eq1}).  While solving Lax pair equations  if we choose plane wave solution as seed solution then we can obtain $N$th order breather solution from the solution formula (\ref{eq11}).  By applying a particular limiting process on the $N$th order breather solution we can derive $N$th order RW solutions of GNLS Eq. (\ref{eq1}).
\par In the following, we present the procedure of deriving positon solutions of GNLS equation.  Here, we impose a limit on the eigenvalues in such a way that, $\lambda_i\to\lambda_1+\epsilon,\;\;i=2,3,...,N$, where $\epsilon$ is a small real parameter.  Evaluating higher order Taylor expansions of eigenfunctions at $\epsilon$ we can obtain positon solutions or degenerate soliton solutions of GNLS equation.  Hence, to determine the degenerate soliton solutions of GNLS equation we transform the solution formula (\ref{eq11}) into
\begin{eqnarray}
	q[N]=q-2i\frac{|D_{1}^{'[N]}|}{|D_{2}^{'[N]}|},
	\label{eq14}
\end{eqnarray}
where
\begin{eqnarray}
	D_{1}^{'[N]}=\left[\frac{\partial^{N_i-1}}{\partial\epsilon^{N_i-1}}|_{\epsilon=0}(D_{1}^{[N]})_{ij}(\lambda_1+\epsilon)\right]_{2N\times2N},\\
	D_{2}^{'[N]}=\left[\frac{\partial^{N_i-1}}{\partial\epsilon^{N_i-1}}|_{\epsilon=0}(D_{2}^{[N]})_{ij}(\lambda_1+\epsilon)\right]_{2N\times2N},
\end{eqnarray}
here $N_i=\left[\frac{i+1}{2}\right]$ with $[i]$ denotes the floor function of $i$.  In the next section, we shall derive explicit expression for positon solutions of GNLS Eq. (\ref{eq1}) from the solution formula (\ref{eq14}).
%%%%%%%%%%%%%%%%%%%%%%%%%%%%%%%%%%%%%%%%%%%%%%%%%%%%%%%%%%%
\section{Smooth positon solutions of GNLS equation}  
\subsection{Second order smooth positon solution}
\begin{figure}
	\includegraphics[width=\linewidth]{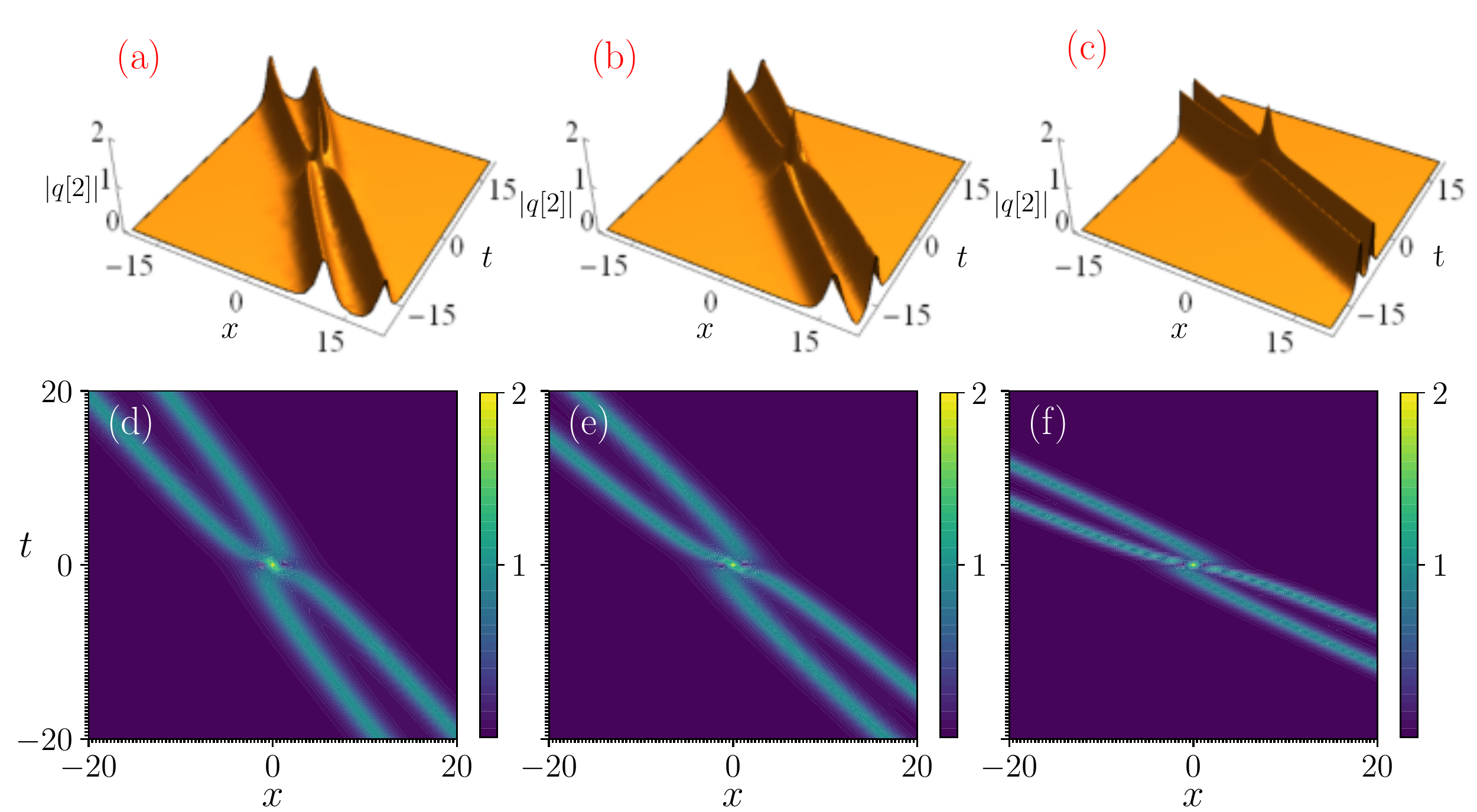}
	\caption{Second order smooth-positon solution for the parameter values $\lambda_1=0.2+0.5i$, (a) $\nu=0$, (b) $\nu=0.2$, (c) $\nu=1$.  Figs. (d)-(f) are the corresponding contour plots of Figs. (a)-(c).}
\end{figure}
To obtain smooth positon solutions, first we solve the Lax pair equations (\ref{eq8}) with zero seed solution, that is $q=0$ with $\lambda=\lambda_1$.  The corresponding eigenfunctions read
\begin{eqnarray}
	\psi_1=\exp(-i\lambda_1x-2i\lambda_{1}^2t+8i\lambda_{1}^4\nu t),\;\
	\phi_1=\exp(i\lambda_1x+2i\lambda_{1}^2t-8i\lambda_{1}^4\nu t).
	\label{eq15}
\end{eqnarray} 
Substituting the above eigenfunctions in the solution formula (\ref{eq11}) with $N=1$ we can obtain one soliton solution.  To determine the positon solution we set $N=2$ in (\ref{eq11}) which in turn provides
\begin{eqnarray}
	q[2]=q-2i\frac{|D_{1}^{[2]}|}{|D_{2}^{[2]}|};
	\label{eq16}
\end{eqnarray}
with
\begin{eqnarray}
	D_{1}^{[2]}=\begin{pmatrix}\psi_1&\phi_1&\lambda_{1}\psi_1&\lambda_{1}^{2}\psi_1\\-\phi_1^*&\psi_1^{*}&-\lambda_1^*\phi_1^*&-\lambda_1^{*2}\phi_1^*\\\psi_2&\phi_2&\lambda_2\psi_2&\lambda_{2}^2\psi_2\\-\phi_2^*&\psi_{2}^*&-\lambda_2^*\phi_2^*&-\lambda_2^{*2}\phi_{2}^*\end{pmatrix},\;\;
	D_{2}^{[2]}=\begin{pmatrix}\psi_1&\phi_1&\lambda_{1}\psi_1&\lambda_{1}\phi_1\\-\phi_1^*&\psi_1^{*}&-\lambda_1^*\phi_1^*&\lambda_1^{*}\psi_1^*\\\psi_2&\phi_2&\lambda_2\psi_2&\lambda_{2}\phi_2\\-\phi_2^*&\psi_{2}^*&-\lambda_2^*\phi_2^*&\lambda_2^{*}\psi_{2}^*\end{pmatrix}. 
	\label{eq17}
\end{eqnarray}
\par Now applying the limit $\lambda_2\to\lambda_1+\epsilon$ and simplifying (\ref{eq16}) with $q=0$ we obtain second order smooth positon solution of GNLS Eq. (\ref{eq1}) in the form
\begin{eqnarray}
	q[2] = A_1/B_1;
	\label{eq18}
\end{eqnarray}
where
\begin{eqnarray}
	A_1&=&4e^{2i(\lambda_1x+2t(\lambda_{1}^2+4\lambda_1^4\nu+8\lambda_{1}^{*4}\nu))}(\lambda_1-\lambda_1^*)(-i+(\lambda_1^*-\lambda_1)x+4t\lambda_1(\lambda_1-\lambda_1^*)(-1+8\lambda_1^2\nu))\nonumber\\&&-4e^{2i(\lambda_1^*x+2t(\lambda_{1}^{*2}+8\lambda_1^4\nu+4\lambda_{1}^{*4}\nu))}(\lambda_1^*-\lambda_1)(-i+(\lambda_1-\lambda_1^*)x-4t\lambda_1^*(\lambda_1-\lambda_1^*)(-1+8\lambda_1^{*2}\nu)),\nonumber\\
	B_1&=&e^{4i(\lambda_1^*x+2t(\lambda_1^{*2}+4\nu\lambda_1^4))}+e^{4i(\lambda_1x+2t(\lambda_1^{2}+4\nu\lambda_1^{*4}))}-2e^{2i(\lambda_1+\lambda_1^*)x+4it(\lambda_1^2+\lambda_1^{*2}+4\nu(\lambda_1^4+\lambda_1^{*4}))}\nonumber\\&&\times(-1+2x^2(\lambda_1-\lambda_1^*)^2+32t^2\lambda_1(\lambda_1-\lambda_1^*)^2\lambda_1^*(-1+8\lambda_1^2\nu)(-1+8\lambda_1^{*2}\nu)\nonumber\\&&-8tx(\lambda_1-\lambda_1^*)^2(\lambda_1+\lambda_1^*)(-1+8\lambda_1^2\nu-8\lambda_1\lambda_1^*\nu+8\lambda_1^{*2}\nu)).	
\end{eqnarray}
\par The second order smooth positon solution is plotted in Fig. 1. Let us analyze how this solution is affected by the higher order nonlinear terms.  To do so we first plot the solution (\ref{eq18}) with $\nu=0$ which is shown in Fig.1 (a).  For $\nu=0$, the solution (\ref{eq18}) becomes positon solution of NLS equation.  Hence Fig.1(a) shows second order smooth positon plot of NLS equation.  Upon increasing the value of the parameter $\nu$ we observe that the positons are being compressed that is the width of both the positons or degenerate solitons decreased.  And the distance between two positons are also get minimized.  Moreover, the direction of positons are also changed.  From Fig. 1(c) we can see that for $\nu=1$, the positons are compressed much further and placed very close to each other.  Hence higher order nonlinear terms make a high compression effect on smooth positons and changes their directions.
\par  Recently it has been shown that the mass and energy of second order positon or degenerate soliton solutions of NLS type equations are twice the mass and energy of one soliton solution \cite{Fringe}.  We have calculated mass and energy of one soliton and two degenerate solution of GNLS equation at their maximum points and we obtained the same expressions for mass and energy as given in \cite{Fringe}.  Hence we do not present them here.
\subsection{Third order smooth positons}
Now we assume $N=3$ in the solution formula (\ref{eq11}) so that we obtain
\begin{eqnarray}
	q[3]=q-2i\frac{|D_{1}^{[3]}|}{|D_{2}^{[3]}|};
	\label{eq19}
\end{eqnarray}
with
\begin{eqnarray}
	D_{1}^{[3]}&=&\begin{pmatrix}\psi_1&\phi_1&\lambda_1\psi_1&\lambda_1\phi_1&\lambda_1^2\psi_1&\lambda_{1}^{3}\psi_1\\-\phi_1^*&\psi_1^{*}&-\lambda_{1}^*\phi_1^*&\lambda_{1}^*\psi_1^*&-\lambda_{1}^{*2}\phi_1^*&-\lambda_{1}^{*3}\phi_1^*\\\psi_2&\phi_2&\lambda_2\psi_2&\lambda_2\phi_2&\lambda_2^2\psi_2&\lambda_2^3\psi_2\\-\phi_2^*&\psi_2^*&-\lambda_2^*\phi_2^*&\lambda_2^*\psi_{2}^{*}&-\lambda_{2}^{*2}\phi_2^*&-\lambda_{2}^{*3}\phi_2^*\\\psi_3&\phi_3&\lambda_3\psi_3&\lambda_3\phi_{3}&\lambda_{3}^2\psi_3&\lambda_{3}^{3}\psi_3\\-\phi_3^*&\psi_3^*&-\lambda_3^*\phi_3^*&\lambda_3^*\psi_3^*&-\lambda_3^{*2}\phi_3^*&-\lambda_3^{*3}\phi_3^*\end{pmatrix},\nonumber\\
	D_{2}^{[3]}&=&\begin{pmatrix}\psi_1&\phi_1&\lambda_1\psi_1&\lambda_1\phi_1&\lambda_1^2\psi_1&\lambda_{1}^{2}\phi_1\\-\phi_1^*&\psi_1^{*}&-\lambda_{1}^*\phi_1^*&\lambda_{1}^*\psi_1^*&-\lambda_{1}^{*2}\phi_1^*&\lambda_{1}^{*2}\psi_1^*\\\psi_2&\phi_2&\lambda_2\psi_2&\lambda_2\phi_2&\lambda_2^2\psi_2&\lambda_2^2\phi_2\\-\phi_2^*&\psi_2^*&-\lambda_2^*\phi_2^*&\lambda_2^*\psi_{2}^{*}&-\lambda_{2}^{*2}\phi_2^*&\lambda_{2}^{*2}\psi_2^*\\\psi_3&\phi_3&\lambda_3\psi_3&\lambda_3\phi_{3}&\lambda_{3}^2\psi_3&\lambda_{3}^{2}\phi_3\\-\phi_3^*&\psi_3^*&-\lambda_3^*\phi_3^*&\lambda_3^*\psi_3^*&-\lambda_3^{*2}\phi_3^*&\lambda_3^{*2}\psi_3^*\end{pmatrix},	 
	\label{eq20}
\end{eqnarray}
\begin{figure}[!ht]
	\includegraphics[width=\linewidth]{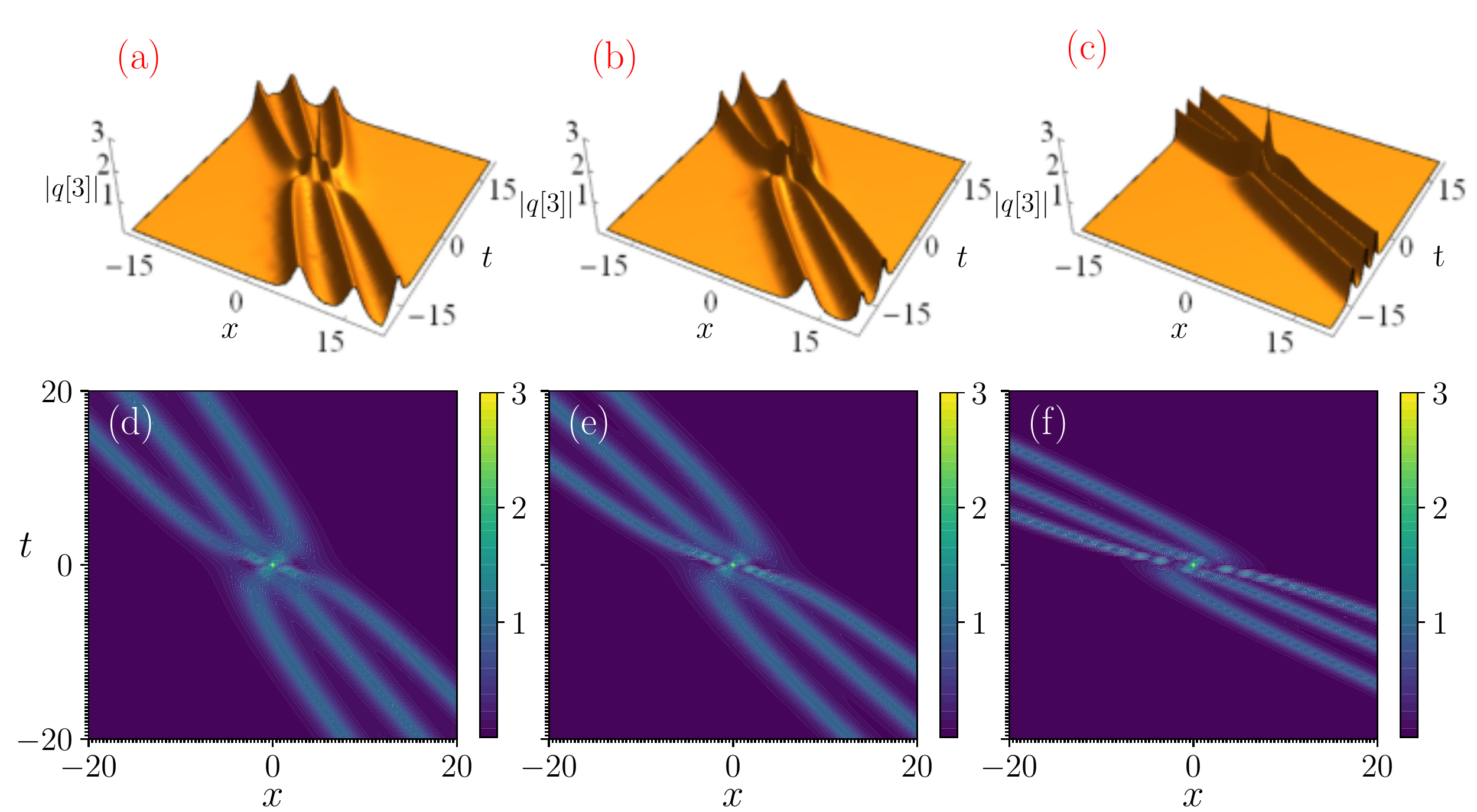}
	\caption{Third order smooth-positon solution for the parameter values $\lambda_1=0.2+0.5i$, (a) $\nu=0$, (b) $\nu=0.2$, (c) $\nu=1$.  Figs. (d)-(f) are the corresponding contour plots of Figs. (a)-(c).}
\end{figure}
\begin{figure}[!ht]
	\includegraphics[width=\linewidth]{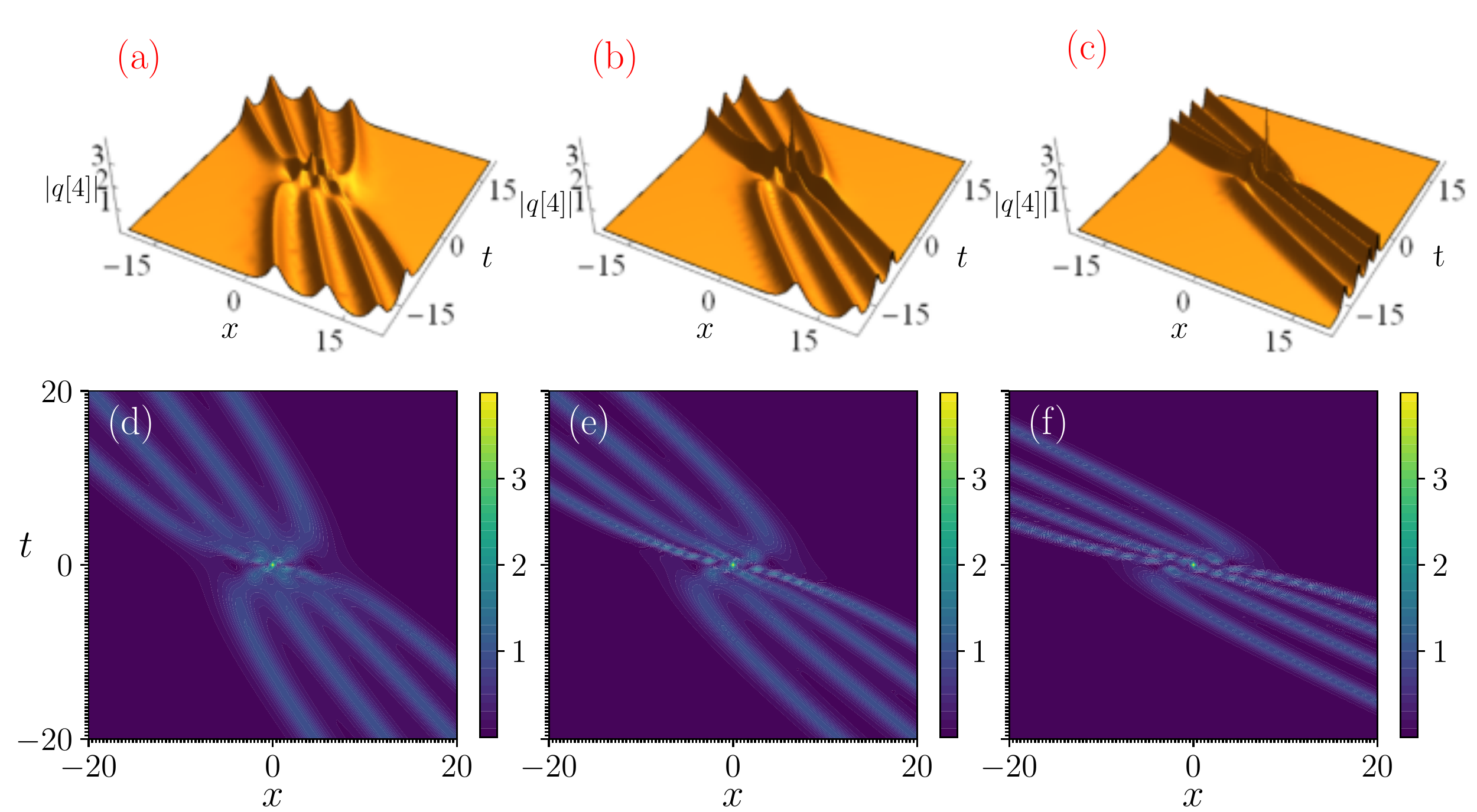}
	\caption{Fourth order smooth-positon solution for the parameter values $\lambda_1=0.2+0.5i$, (a) $\nu=0$, (b) $\nu=0.2$, (c) $\nu=1$.  Figs. (d)-(f) are the corresponding contour plots of Figs. (a)-(c).}
\end{figure}
\par By taking the limit $\lambda_i\to\lambda_1+\epsilon,\;\;i=2,3,$ with $q=0$ to the solution (\ref{eq19}) and simplifying the resultant equation, we obtain third order positon solution.  Since it is a very lengthy expression, we present its exact form in Appendix A.  We plot the solution (\ref{AppAeq}) in Fig.2.   Now we analyze the effect of higher order terms on the third order positon solution of GNLS Eq. (\ref{eq1}).  For $\nu=0$ Fig. 2(a) shows third order positon solution of NLS equation.  Now we increase the value of the parameter $\nu$ to 0.2.  As one notices the width of the positons are compressed and they come close to each other.  The directions of the positons also change.  When $\nu=1$, the positons are well compressed and the distance between them becomes very small.  Hence, as in the previous case, for higher values of $\nu$ positons get compressed and their directions change.  Figures 2(b) and 2(c) reveal the compression effect with respect to the parameter $\nu$ on third order positon solution of GNLS equation (\ref{eq1}).  
\par Fourth order positon solution of GNLS equation can also be derived from the solution formula (\ref{eq14}) with $N=4$.  Since the fourth order solution is very lengthy we do not present the explicit form of it here and present only the plot of it.  The parameter $\nu$ exhibits the same compression effect on fourth order positon solution as well.  Figure 3(a) shows the fourth order positon solution of NLS equation and the compression effect of the parameter $\nu$ on the fourth order positon solution of the GNLS equation (\ref{eq1}) can be confirmed from Figs. 3(b) and 3(c). 
%%%%%%%%%%%%%%%%%%%%%%%%%%%%%%%%%%%%%%%%%%%%
\section{B-P solutions of GNLS equation (\ref{eq1})}
\subsection{Second order B-P solution}
\begin{figure}
	\includegraphics[width=\linewidth]{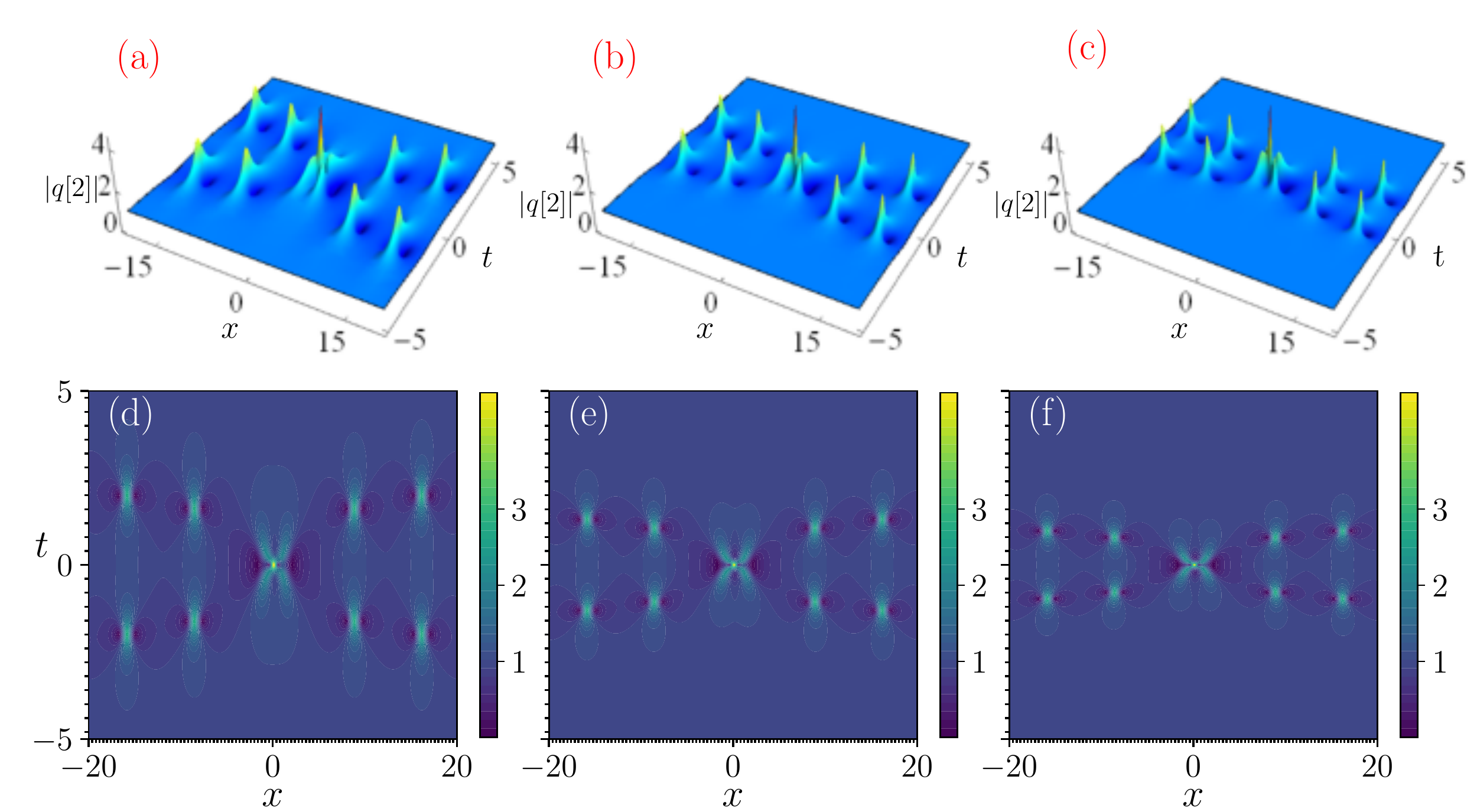}
	\caption{Second order breather-positon solution of GNLS equation for the parameter values $a=1$, $\lambda_1=0.9i$, $s_{0r}=s_{0i}=0$, (a) $\nu=0$, (b) $\nu=0.1$ and (c) $\nu=0.2$.  Figs. (d)-(f) are the corresponding contour plots of Figs. (a)-(c).}
\end{figure}
\begin{figure}
	\includegraphics[width=\linewidth]{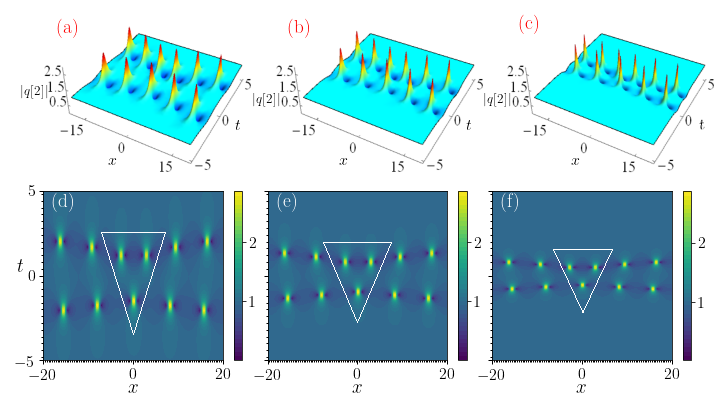}
	\caption{Triplet structure of second order breather-positon solution of GNLS equation for the parameter values $a=1$, $\lambda_1=0.9i$, $s_{0r}=25$ and $s_{0i}=0$, (a) $\nu=0$, (b) $\nu=0.1$ and (c) $\nu=0.2$.  Figs. (d)-(f) are the corresponding contour plots of Figs. (a)-(c).}
\end{figure}
To determine the B-P solutions of GNLS equation one should solve the Lax pair equations (\ref{eq8}) with plane wave as seed solution, that is $q=ae^{2ia^2(1+3a^2\nu)t}$, where $a$ is a real constant.  By solving them with the chosen form of seed solution we obtain the following expressions for the eigenfunctions, that is
\begin{eqnarray}
	\psi_1&=&(c_1e^{A}+c_2e^{-A})e^{ia^2(1+3a^2\nu)t},\
	\phi_1=(c_3e^{A}+c_4e^{-A})e^{-ia^2(1+3a^2\nu)t},\nonumber\\
	A&=&i\sqrt{\lambda_1^2+a^2}(x+(2i(\lambda_1+2a^2\lambda_1\nu-4\lambda_1^3\nu))),\nonumber \\
	c_1&=&\frac{ac_3}{i\sqrt{\lambda_1^2+a^2}+i\lambda_1},\qquad \quad c_2=\frac{ac_4}{-i\sqrt{\lambda_1^2+a^2}+i\lambda_1},\nonumber\\
	c_3&=&\frac{ia+2(\lambda_1+\sqrt{a^2+\lambda_1^2})}{\sqrt{a^2+\lambda_1^2}},\
	c_4=\frac{-ia-2(\lambda_1-\sqrt{a^2+\lambda_1^2})}{\sqrt{a^2+\lambda_1^2}}.\label{eq21}
\end{eqnarray}
\par By substituting the above eigenfunctions in (\ref{eq16}) and applying the limit $\lambda_2\to\lambda_1+\epsilon$ we obtain second order B-P solution or degenerate breather solution of GNLS Eq. (\ref{eq1}).  We present the explicit form of second order B-P solution in Appendix B.  Now we analyze the effect of higher order nonlinear terms on B-P solutions as we did in the case of smooth positons case.  Figure 4(a) is plotted for $\nu=0$ which shows second order B-P solution of NLS equation.  We can visualize the central region of B-P solution exhibits a structure similar as that of second order RW solution, that is a single highest crest in the center and four sub-crests around it.  When we increase the value of $\nu$ to 0.1, the width of B-P pulses and the distance between two B-Ps get decreased.  The direction of B-Ps are also affected and the width of the central region of B-P solution is decreased.  By increasing the value of the parameter $\nu$ to 0.2 we obtain well compressed B-P pulses and here also we observe that the distance between two B-Ps decrease.  The central region of B-Ps is also highly affected by the higher order nonlinear terms which can be clearly seen from the Figs. 4(b) and 4(c).  When we introduce an arbitrary constant in the exponential function $A$ given in Eq. (\ref{eq21}), that is $A=i\sqrt{\lambda_1^2+a^2}(x+(2i(\lambda_1+2a^2\lambda_1\nu-4\lambda_1^3\nu))+s_0\epsilon)$, with $s_0=s_{0r}+is_{0i}$ we can obtain an interesting central region in the second order B-P solution.  For example, if we choose $s_{0r}=s_{0i}=25$, the central region of second order B-P solution, that is the second order RW like structure splits into three single breather pulses.  This structure is known as triplet of RWs.  A similar behaviour is also exhibited by second order RW solution.  For higher values of $\nu$, we come across greater compression in the triplet pattern as shown in Fig. 5.  

\par In Ref. \cite{porsezian1}, the authors have constructed higher order breather solutions of GNLS equation.  They have shown that the period of breather solutions is highly affected by the higher order nonlinear terms.  They have also shown that when we increase the value of the parameter $\nu$, the period of breather also increases.  But differing from that, here our results reveal that the period of B-P solution is not affected by the parameter $\nu$.  When we increase the value of the parameter $\nu$, B-P pulses get compressed but the number of pulse do not increase.  This is the major difference which one can notice between breather and B-P solutions of GNLS equation.  To the best of our knowledge, this particular behaviour of B-P solution has not been mentioned in the literature.  
\subsection{Third order B-P solution}
\begin{figure}
	\includegraphics[width=\linewidth]{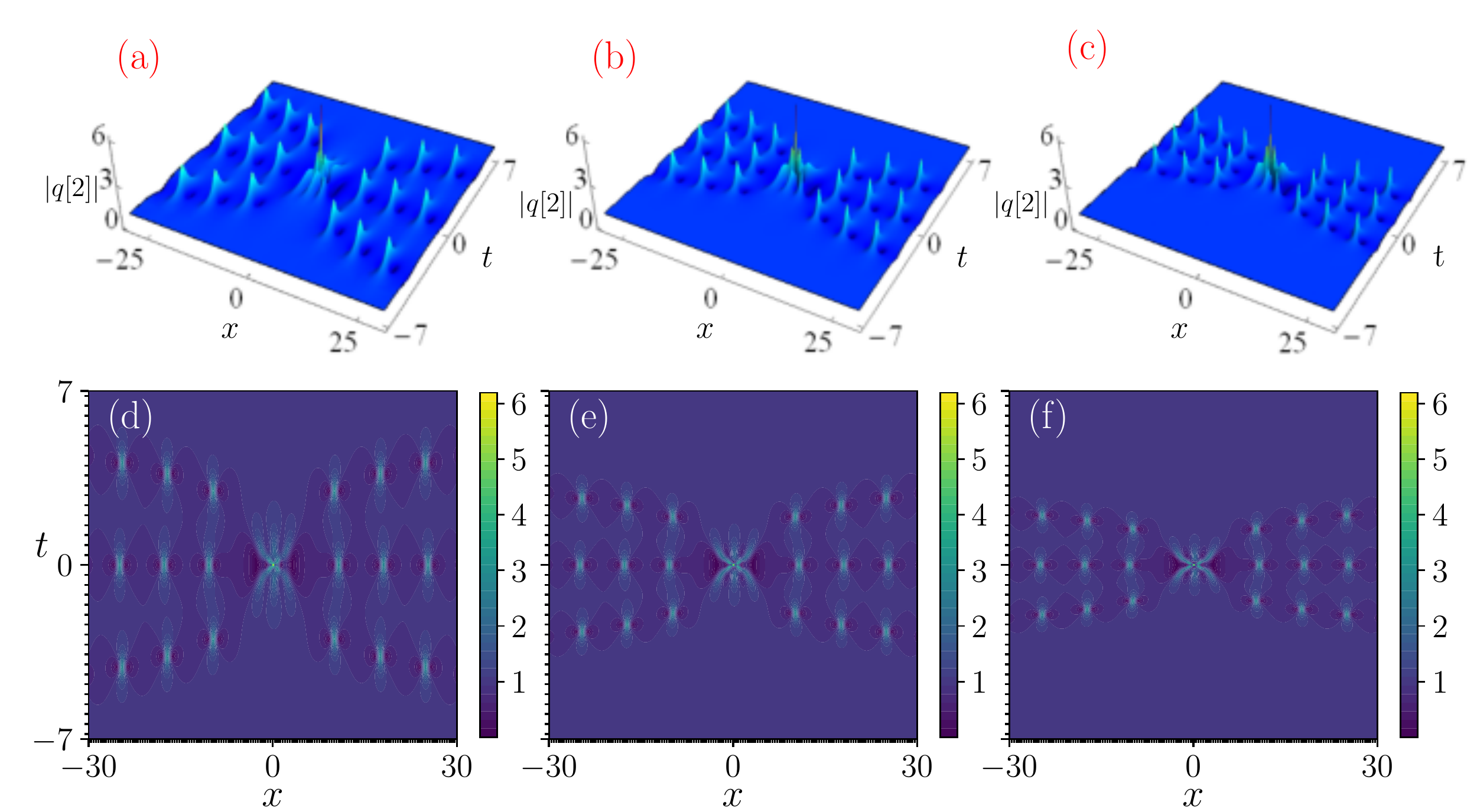}
	\caption{Third order breather-positon solution of GNLS equation for the parameter values $a=1$, $\lambda_1=0.9i$, $s_{0r}=s_{0i}=s_{1r}=s_{1i}=0$, (a) $\nu=0$, (b) $\nu=0.1$ and (c) $\nu=0.2$.  Figs. (d)-(f) are the corresponding contour plots of Figs. (a)-(c).}
\end{figure}
\begin{figure}
	\includegraphics[width=\linewidth]{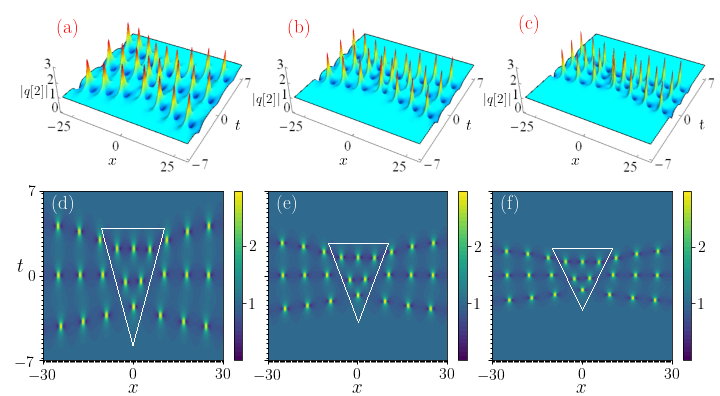}
	\caption{Triangular structure of third order breather-positon solution of GNLS equation for the parameter values $a=1$, $\lambda_1=0.9i$, $s_{0r}=25,\ s_{0i}=0, \ s_{1r}=25, \ s_{1i}=25$, (a) $\nu=0$, (b) $\nu=0.1$ and (c) $\nu=0.2$.  Figs. (d)-(f) are the corresponding contour plots of Figs. (a)-(c).}
\end{figure}
\begin{figure}
	\includegraphics[width=\linewidth]{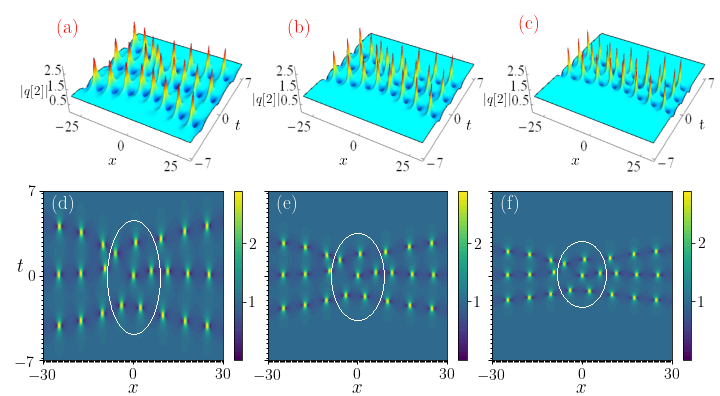}
	\caption{Circular structure of third order breather-positon solution of GNLS equation for the parameter values $a=1$, $\lambda_1=0.9i$, $s_{0r}=0,\ s_{0i}=0, \ s_{1r}=400, \ s_{1i}=400$, (a) $\nu=0$, (b) $\nu=0.1$ and (c) $\nu=0.2$.  Figs. (d)-(f) are the corresponding contour plots of Figs. (a)-(c).}
\end{figure}
To obtain the third order B-P solution of GNLS Eq. (\ref{eq1}), we substitute the eigenfunctions (\ref{eq21}) with the limit $\lambda_i\to\lambda_1+\epsilon,\;\;i=1,2,3$.  Since the third order B-P solution is very lengthy, we have only plotted it in Fig. 6.  The central region of third order B-P does exhibit a similar structure as that of third order rogue wave.  Higher compression effect happens in the third order B-P solution for higher values of $\nu$ as in the previous cases which are demonstrated in Figs. 6(b) and 6(c).  Now we introduce an arbitrary constant in the eigenfunctions that is $A=i\sqrt{\lambda_1^2+a^2}(x+(2i(\lambda_1+2a^2\lambda_1\nu-4\lambda_1^3\nu))+s_0\epsilon+s_1\epsilon^2)$, where $s_1=s_{1r}+is_{1i}$.  By choosing $s_0=0$ and $s_{1r}=s_{1i}=400$, the central region of third order B-P solution splits into six B-P pulses which forms a triangular form as shown in Fig. 7.  When we take $s_1=0$ and $s_{or}=s_{oi}=25$ the central region shows a circular form of six B-P pulses as shown in Fig. 8.  These triangular and circular patterns of central region of the third order B-P are also highly compressed for higher values of $\nu$ which can be clearly seen from Figs.7 and 8.  
\section{Conclusions}
In this paper, we have derived the degenerate solutions of GNLS Eq. (\ref{eq1}) using GDT method.  By choosing the vacuum seed solution in GDT, we have generated smooth positon solutions of GNLS equation.  We have shown that width of smooth positon solutions and the distance between two positons are highly compressed by the higher order nonlinear parameter $\nu$.  The parameter $\nu$ also changes the direction of the smooth positons.  We have also constructed the B-P solutions of GNLS equation using plane wave solution as seed solution in the GDT method.  The higher order nonlinear terms make similar compression effect on B-P solutions.  The width and the distance between two B-Ps are highly affected by the parameter $\nu$.  In the literature, it has been shown that the period of breather solution increases when one increases the value of $\nu$.  In this work we have demonstrated that the period of B-P solution is not affected by the parameter $\nu$.   We believe that various nonlinear wave solutions discussed in this paper such as degenerate solitons, breathers and their compression effects due to higher order nonlinear terms will be useful to the optics and ferromagnetic research community.  
%%%%%%%%%%%%%%%%%%%%%%%%%%%%%%%%%%%%%%%%%%%%%%%%%%%%%%%%%%%%%%%%%%%%%%%%%%%%%%%%%%%%%%%%%%%%%%%%%%%%%%%%%%%%%
\appendix
\section*{Appendix A: \label{AppA}Third order smooth positon solution of GNLS equation}
Third order smooth positon solution of GNLS equation is given by,
\begin{eqnarray}
	q[3]=\frac{A_2}{B_2},\label{AppAeq}
\end{eqnarray}
\begin{eqnarray}
	A_2&=&A_{21}e^{i(x(\lambda_1-3\lambda_1^*)+2t(\lambda_1^2-3\lambda_1^{*2}-4\lambda_1^4\nu+12\lambda_1^{*4}\nu))}+A_{22}e^{i(x(-3\lambda_1+\lambda_1^*)+2t(\lambda_1^{*2}-3\lambda_1^{2}-4\lambda_1^{*4}\nu+12\lambda_1^{4}\nu))}\nonumber\\&&+A_{23}e^{-i(x(\lambda_1+\lambda_1^*)+2t(\lambda_1^2+\lambda_1^{*2}-4\lambda_1^4\nu-4\lambda_1^{*4}\nu))},\nonumber\\
	A_{21}&=&-4(\lambda_1-\lambda_1^*)\big(-3+2x^2(\lambda_1-\lambda_1^*)^2+32t^2\lambda_1^2(\lambda_1-\lambda_1^*)^2(1-8\lambda_1^2\nu)^2-4it(\lambda_1-\lambda_1^*)(-7\lambda_1+\lambda_1^*\nonumber\\&&+72\lambda_1^3\nu-24\lambda_1^2\lambda_1^*\nu)-2x(\lambda_1-\lambda_1^*)(-3i+8t\lambda_1(\lambda_1-\lambda_1^*)(-1+8\lambda_1^2\nu))\big),\nonumber\\
	A_{22}&=&-4(\lambda_1-\lambda_1^*)\big(-3+2x^2(\lambda_1-\lambda_1^*)^2+32t^2\lambda_1^{*2}(\lambda_1-\lambda_1^*)^2(1-8\lambda_1^{*2}\nu)^2-4it(\lambda_1-\lambda_1^*)(7\lambda_1^*-\lambda_1\nonumber\\&&-72\lambda_1^{*3}\nu+24\lambda_1^{*2}\lambda_1\nu)-2x(\lambda_1-\lambda_1^*)(3i+8t\lambda_1^*(\lambda_1-\lambda_1^*)(-1+8\lambda_1^{*2}\nu))\big),\nonumber\\
	A_{23}&=&-8(\lambda_1-\lambda_1^*)\bigg(-3-4x^2(\lambda_1-\lambda_1^*)^2+2x^4(\lambda_1-\lambda_1^*)^4+512t^4\lambda_1^2(\lambda_1-\lambda_1^*)^4\lambda_1^{*2}(1-8\lambda_1^2\nu)^2\nonumber\\&&(1-8\lambda_1^{*2}\nu)^2-16t(\lambda_1-\lambda_1^*)^2(x^3(\lambda_1-\lambda_1^*)^2(\lambda_1+\lambda_1^*)(-1+8\lambda_1^2\nu-8\lambda_1\lambda_1^*\nu+8\lambda_1^{*2}\nu)\nonumber\\&&+i(-1+9\lambda_1^2\nu+6\lambda_1\lambda_1^*\nu+9\lambda_1^{*2}\nu)-ix^2(\lambda_1-\lambda_1^*)^2(-1+10\lambda_1^2\nu+4\lambda_1\lambda_1^*\nu+10\lambda_1^{*2}\nu)\nonumber\\&&-x(\lambda_1+\lambda_1^*)(-1+14\lambda_1^2\nu-20\lambda_1\lambda_1^*\nu+14\lambda_1^{*2}\nu))+8t^2(\lambda_1-\lambda_1^*)^2(256x^2\lambda_1^8\nu^2-256x\lambda_1^7(i\nonumber\\&&
	+2x\lambda_1^*)\nu^2+32x\lambda_1^5\nu(3i-24i\lambda_1^{*2}\nu+32x\lambda_1^{*3}\nu)+64\lambda_1^6\nu(-2\nu+4ix\lambda_1^*\nu+x^2(-1+4\lambda_1^{*2}\nu))\nonumber\\&&+\lambda_1^2(-1-48\lambda_1^{*2}\nu-192\lambda_1^{*4}\nu^2+8x^2\lambda_1^{*2}(-3+24\lambda_1^{*2}\nu+32\lambda_1^{*4}\nu^2)-8ix\lambda_1^*(-1+8\lambda_1^{*2}\nu\nonumber\\&&+96\lambda_1^{*4}\nu^2))-4\lambda_1^4(2\nu(-5+24\lambda_1^{*2}\nu)-8ix\lambda_1^*\nu(-1+24\lambda_1^{*2}\nu)+x^2(-1-48\lambda_1^{*2}\nu+512\lambda_1^{*4}\nu^2))\nonumber\\&&+\lambda_1^{*2}(-1+40\lambda_1^{*2}\nu-128\lambda_1^{*4}\nu^2+4x^2\lambda_1^{*2}(1-8\lambda_1^{*2}\nu)^2-8ix(\lambda_1^*-12\lambda_1^{*3}\nu+32\lambda_1^{*5}\nu^2))\nonumber\\&&+8\lambda_1^3(2\lambda_1^*\nu(3+8\lambda_1^{*2}\nu)+ix(-1-8\lambda_1^{*2}\nu+96\lambda_1^{*4}\nu^2)+x^2(\lambda_1^*-32\lambda_1^{*3}\nu+128\lambda_1^{*5}\nu^2))\nonumber\\&&+2\lambda_1\lambda_1^*(-3+24\lambda_1^{*2}\nu+4ix(\lambda_1^-4\lambda_1^{*3}\nu+32\lambda_1^{*5}\nu^2)+x^2(4\lambda_1^{*2}-256\lambda_1^{*6}\nu^2)))\nonumber
\end{eqnarray}
\begin{eqnarray}&&-64t^3(\lambda_1-\lambda_1^*)^4\big(i\lambda_1^{*2}(1-8\lambda_1^{*2}\nu)^2-2\lambda_1\lambda_1^*(-i+2x\lambda_1^*)(1-8\lambda_1^{*2}\nu)^2-128i\lambda_1^5\lambda_1^*\nu^2\nonumber\\&&\times(-1+8\lambda_1^{*2}\nu)+32\lambda_1^3\lambda_1^*\nu(-1+8\lambda_1^{*2}\nu)(i-4i\lambda_1^{*2}\nu+x\lambda_1^*(-1+8\lambda_1^{*2}\nu))+64\lambda_1^6\nu^2(i-24i\lambda_1^{*2}\nu\nonumber\\&&+4x\lambda_1^*(-1+8\lambda_1^{*2}\nu))+16\lambda_1^4\nu(x(4\lambda_1^*-32\lambda_1^{*3}\nu)-i(1-32\lambda_1^{*2}\nu+64\lambda_1^{*4}\nu^2))\nonumber\\&&+\lambda_1^2(4x\lambda_1^*(-1+8\lambda_1^{*2}\nu)-i(-1+64\lambda_1^{*2}\nu-512\lambda_1^{*4}\nu^2+1536\lambda_1^{*6}\nu^3))\big)\bigg),\nonumber\\
	B_2&=&4e^{3i(x(-\lambda_1+\lambda_1^*)+2t(-\lambda_1^2+\lambda_1^{*2}+4\lambda_1^4\nu-4\lambda_1^{*4}\nu))}+4e^{3i(x(\lambda_1-\lambda_1^*)+2t(\lambda_1^2-\lambda_1^{*2}-4\lambda_1^4\nu+4\lambda_1^{*4}\nu))}\nonumber\\&&+4B_{21}e^{-i(x(-\lambda_1+\lambda_1^*)+2t(-\lambda_1^2+\lambda_1^{*2}+4\lambda_1^4\nu-4\lambda_1^{*4}\nu))}+4B_{22}e^{-i(x(\lambda_1-\lambda_1^*)+2t(\lambda_1^2-\lambda_1^{*2}-4\lambda_1^4\nu+4\lambda_1^{*4}\nu))},\nonumber\\
	B_{21}&=&(3+4x^4(\lambda_1-\lambda_1^*)^4+1024t^4\lambda_1^2(\lambda_1-\lambda_1^*)^4\lambda_1^{*2}(1-8\lambda_1^2\nu)^2(1-8\lambda_1^{*2}\nu)^2-8x^3(\lambda_1-\lambda_1^*)^3(-i\nonumber\\&&+4t(\lambda_1^2-\lambda_1^{*2})(-1+8\lambda_1^2\nu-8\lambda_1\lambda_1^*\nu+8\lambda_1^{*2}\nu))+16t^2(\lambda_1-\lambda_1^*)^2(3\lambda_1^{*2}-40\lambda_1^{*4}\nu+16\lambda_1^3\lambda_1^*\nu(13\nonumber\\&&-168\lambda_1^{*2}\nu)+40\lambda_1^4\nu(-1+24\lambda_1^{*2}\nu)+2\lambda_1\lambda_1^*(-9+104\lambda_1{^*2}\nu)+3\lambda_1^2(1-48\lambda_1^{*2}\nu+320\lambda_1^{*4}\nu^2))\nonumber\\&&+128it^3(\lambda_1-\lambda_1^*)^3(\lambda_1+\lambda_1^*)(4\lambda_1\lambda_1^*(1-8\lambda_1^{*2}\nu)^2-\lambda_1^{*2}(1-8\lambda_1^{*2}\nu)^2-256\lambda_1^5\lambda_1^*\nu^2(-1+16\lambda_1^{*2}\nu)\nonumber\\&&+64\lambda_1^6\nu^2(-1+24\lambda_1^{*2}\nu)-64\lambda_1^3\lambda_1^*\nu(1-20\lambda_1^{*2}\nu+64\lambda_1^{*4}\nu^2)+16\lambda_1^4\nu(1-40\lambda_1^{*2}\nu+256\lambda_1^{*4}\nu^2)\nonumber\\&&+\lambda_1^2(-1+48\lambda_1^{*2}\nu-640\lambda_1^{*4}\nu^2+1536\lambda_1^{*6}\nu^3))+4x^2(\lambda_1-\lambda_1^*)^2(-3-12it(\lambda_1^2-\lambda_1^{*2})(-1+12\lambda_1^2\nu\nonumber\\&&-16\lambda_1\lambda_1^*\nu+12\lambda_1^{*2}\nu)+16t^2(\lambda_1-\lambda_1^*)^2(\lambda_1^2-16\lambda_1^4\nu+64\lambda_1^6\nu^2+\lambda_1^{*2}(1-8\lambda_1^{*2}\nu)^2+32\lambda_1^3\lambda_1^*\nu(-1\nonumber\\&&+8\lambda_1^{*2}\nu)+\lambda_1(4\lambda_1^*-32\lambda_1^{*3}\nu)))-16tx(\lambda_1-\lambda_1^*)^2(2048t^2\lambda_1^8\lambda_1^*\nu^2(-1+8\lambda_1^{*2}\nu)+\lambda_1^*(3-36\lambda_1^{*2}\nu\nonumber\\&&-32it\lambda_1^{*4}\nu+256it\lambda_1^{*6}\nu^2)-256t\lambda_1^7\nu^2(i+16t\lambda_1^{*2}(-1+8\lambda_1^{*2}\nu))+\lambda_1(3+12\lambda_1^{*2}\nu-32t^2\lambda_1^{*4}\nonumber\\&&\times(1-8\lambda_1^{*2}\nu)^2-8it\lambda_1^{*2}(-3+28\lambda_1^{*2}\nu+32\lambda_1^{*4}\nu^2))+256t\lambda_1^6\lambda_1^*\nu(i\nu+t(2-24\lambda_1^{*2}\nu+64\lambda_1^{*4}\nu^2))\nonumber\\&&+32t\lambda_1^5\nu(i(1+24\lambda_1^{*2}\nu)+8t\lambda_1^{*2}(-3+16\lambda_1^{*2}\nu+64\lambda_1{^*4}\nu^2))+4\lambda_1^2\lambda_1^*(3\nu-2it(3-56\lambda_1^{*2}\nu\nonumber\\&&+96\lambda_1^{*4}\nu^2)+8t^2(\lambda_1^{*2}-24\lambda_1^{*4}\nu+128\lambda_1^{*6}\nu^2))-32t\lambda_1^4\lambda_1^*(i\nu(-7+104\lambda_1^{*2}\nu)+t(1-8\lambda_1^{*2}\nu\nonumber\\&&-128\lambda_1^{*4}\nu^2+1024\lambda_1^{*6}\nu^3))+4\lambda_1^3(-9\nu+16it\lambda_1^{*2}\nu(-7+52\lambda_1^{*2}\nu)+8t^2(\lambda_1^{*2}+8\lambda_1^{*4}\nu\nonumber\\&&-192\lambda_1^{*6}\nu^2+512\lambda_1^{*8}\nu^3)))),\nonumber\\
	%%%%%%%%%%%%%%%%%%%%%%%
	%%%%%%%%%%%%%%%%%%%%%%
	B_{22}&=&(3+4x^4(\lambda_1-\lambda_1^*)^4+1024t^4\lambda_1^2(\lambda_1-\lambda_1^*)^4\lambda_1^{*2}(1-8\lambda_1^2\nu)^2(1-8\lambda_1^{*2}\nu)^2-8x^3(\lambda_1-\lambda_1^*)^3(i\nonumber\\&&+4t(\lambda_1^2-\lambda_1^{*2})(-1+8\lambda_1^2\nu-8\lambda_1\lambda_1^*\nu+8\lambda_1^{*2}\nu))+16t^2(\lambda_1-\lambda_1^*)^2(3\lambda_1^{*2}-40\lambda_1^{*4}\nu+16\lambda_1^3\lambda_1^*\nu(13\nonumber\\&&-168\lambda_1^{*2}\nu)+40\lambda_1^4\nu(-1+24\lambda_1^{*2}\nu)+2\lambda_1\lambda_1^*(-9+104\lambda_1{^*2}\nu)+3\lambda_1^2(1-48\lambda_1^{*2}\nu+320\lambda_1^{*4}\nu^2))\nonumber\\&&-128it^3(\lambda_1-\lambda_1^*)^3(\lambda_1+\lambda_1^*)(4\lambda_1\lambda_1^*(1-8\lambda_1^{*2}\nu)^2-\lambda_1^{*2}(1-8\lambda_1^{*2}\nu)^2-256\lambda_1^5\lambda_1^*\nu^2(-1+16\lambda_1^{*2}\nu)\nonumber\\&&+64\lambda_1^6\nu^2(-1+24\lambda_1^{*2}\nu)-64\lambda_1^3\lambda_1^*\nu(1-20\lambda_1^{*2}\nu+64\lambda_1^{*4}\nu^2)+16\lambda_1^4\nu(1-40\lambda_1^{*2}\nu+256\lambda_1^{*4}\nu^2)\nonumber\\&&+\lambda_1^2(-1+48\lambda_1^{*2}\nu-640\lambda_1^{*4}\nu^2+1536\lambda_1^{*6}\nu^3))+4x^2(\lambda_1-\lambda_1^*)^2(-312it(\lambda_1^2-\lambda_1^{*2})(-1+12\lambda_1^2\nu\nonumber\\&&-16\lambda_1\lambda_1^*\nu+12\lambda_1^{*2}\nu)+16t^2(\lambda_1-\lambda_1^*)^2(\lambda_1^2-16\lambda_1^4\nu+64\lambda_1^6\nu^2+\lambda_1^{*2}(1-8\lambda_1^{*2}\nu)^2+32\lambda_1^3\lambda_1^*\nu(-1\nonumber\\&&+8\lambda_1^{*2}\nu)+\lambda_1(4\lambda_1^*-32\lambda_1^{*3}\nu)))-16tx(\lambda_1-\lambda_1^*)^2(2048t^2\lambda_1^8\lambda_1^*\nu^2(-1+8\lambda_1^{*2}\nu)+\lambda_1^*(3-36\lambda_1^{*2}\nu\nonumber\end{eqnarray}
\begin{eqnarray} &&32it\lambda_1^{*4}\nu-256it\lambda_1^{*6}\nu^2)-256t\lambda_1^7\nu^2(-i+16t\lambda_1^{*2}(-1+8\lambda_1^{*2}\nu))+\lambda_1(3+12\lambda_1^{*2}\nu\nonumber\\&&-32t^2\lambda_1^{*4}(1-8\lambda_1^{*2}\nu)^28it\lambda_1^{*2}(-3+28\lambda_1^{*2}\nu+32\lambda_1^{*4}\nu^2))+256t\lambda_1^6\lambda_1^*\nu(-i\nu+t(2-24\lambda_1^{*2}\nu\nonumber\\&&+64\lambda_1^{*4}\nu^2))+32t\lambda_1^5\nu(-i(1+24\lambda_1^{*2}\nu)+8t\lambda_1^{*2}(-3+16\lambda_1^{*2}\nu+64\lambda_1{^*4}\nu^2))+4\lambda_1^2\lambda_1^*(3\nu\nonumber\\&&+2it(3-56\lambda_1^{*2}\nu+96\lambda_1^{*4}\nu^2)+8t^2(\lambda_1^{*2}-24\lambda_1^{*4}\nu+128\lambda_1^{*6}\nu^2))-32t\lambda_1^4\lambda_1^*(-i\nu(-7+104\lambda_1^{*2}\nu)\nonumber\\&&+t(1-8\lambda_1^{*2}\nu-128\lambda_1^{*4}\nu^2+1024\lambda_1^{*6}\nu^3))+4\lambda_1^3(-9\nu-16it\lambda_1^{*2}\nu(-7+52\lambda_1^{*2}\nu)+8t^2(\lambda_1^{*2}+8\lambda_1^{*4}\nu\nonumber\\&&-192\lambda_1^{*6}\nu^2+512\lambda_1^{*8}\nu^3)))).
\end{eqnarray}
%%%%%%%%%%%%%%%%%%%%%%%%%%
\section*{Appendix B: \label{AppB}Second order B-P solution of GNLS equation}
The second order B-P solution of (\ref{eq1}) is given by
\begin{eqnarray}
	q[2]_{b-p}=ae^{ict}-2i\frac{A_{bp2}}{B_{bp2}},
\end{eqnarray}
where
\begin{eqnarray}
	A_{bp2}&=&\frac{8a}{\delta^5}\eta(5a+4\eta)^2(a^2-2\eta^2-2i\eta\delta)e^{2it(a^2+3a^4\nu)-8t\delta(\eta+2a^2\eta\nu+4\eta^3\nu)}-\frac{8a}{\delta^5}\eta(5a+4\eta)^2(a^2\nonumber\\&&+2i\eta(i\eta+\delta))e^{2t(3ia^4\nu+4\eta\delta(1+4\eta^2\nu)+a^2(i+8\eta\delta\nu))}+\frac{1}{\delta^5}1280ia^8\eta^2\nu te^{2ix\delta+2it(a^2+3a^4\nu)-4t\delta(\eta+2a^2\eta\nu+4\eta^3\nu)}\nonumber\\&&-\frac{1}{\delta^2}1280ia^8\eta^2\nu te^{2ix\delta+2it(a^2+3a^4\nu)+4t\delta(\eta+2a^2\eta\nu+4\eta^3\nu)}-\frac{32a}{\delta^4}\eta^2(5a+4\eta)^2(i+8a^4\nu t\nonumber\\&&+4a^2(t+8t\eta^2\nu)-8t(\eta^2+8\eta^4\nu))+\frac{8\eta}{\delta^5}G_1e^{-2ix\delta+2it(a^2+3a^4\nu)-4t\delta(\eta+2a^2\eta\nu+4\eta^3\nu)}\nonumber\\&&-\frac{8\eta}{\delta^5}G_2e^{6ia^4t\nu+2\delta(ix+2t\eta+8t\eta^3\nu)+2a^2t(i+4\eta\delta\nu)}
	+\frac{8\eta}{\delta^5}G_3e^{-2ix\delta+2it(a^2+3a^4\nu)+4t\delta(\eta+2a^2\eta\nu+4\eta^3\nu)}\nonumber\\&&+\frac{8\eta}{\delta^5}G_4e^{6ia^4t\nu-2\delta(-ix+2t\eta+8t\eta^3\nu)+a^2t(2i-8\eta\delta\nu)},\nonumber\\
	G_1&=&160ia^8t\eta\nu-8a^7t\eta(-56i\eta+15\delta)\nu+16a^6t\eta(5i+46i\eta^2\nu+4\eta\delta\nu)+4a^2\eta^2(-56it\eta^3+7i\delta\nonumber\\&&+\eta^2(-6x+16t\delta)+\eta(2+4ix\delta)-448it\eta^5\nu+128t\eta^4\delta\nu)+2a^3\eta(-336it\eta^3+25i\delta+\eta^2(-56x\nonumber\\&&+172t\delta)+3\eta(4-5ix\delta)-2688it\eta^5\nu+1376t\eta^4\delta\nu)-32\eta^4(-i\eta+\delta)(-i-2ix\eta+8t(\eta^2+8\eta^4))\nonumber\\&&-16a\eta^3(-i\eta+\delta)(-2i-7ix\eta+2t(\eta^2+8\eta^4\nu))+a^5(-15+4it\eta(56\eta+15i\delta+336\eta^3\nu\nonumber\\&&+8i\eta^2\delta\nu))+4a^4(-10x\eta^2+5i\delta+4t\eta^2(-7i\eta+2\delta-72i\eta^3\nu+32\eta^2\delta\nu)),\nonumber\\
	G_2&=&8a^7t\eta(56i\eta+15\delta)\nu+16ia^6t\eta(5+46\eta^2\nu+4i\eta\delta\nu)-4a^2\eta^2(56it\eta^3+7i\delta+2\eta^2(3x+8\delta t)\nonumber\\&&+\eta(-2+4ix\delta)+448it\eta^5\nu+128t\eta^4\delta\nu)-2a^3\eta(336it\eta^3+25i\delta+4\eta^2(14x+43t\delta)\nonumber\\\end{eqnarray}
\begin{eqnarray}&&+\eta(-12-15ix\delta)+2688it\eta^5\nu+1376t\eta^4\delta\nu)+32\eta^4(i\eta+\delta)(-i-2ix\eta+8t(\eta^2+8\eta^4\nu))\nonumber\\&&+16a\eta^3(i\eta+\delta)(-2i-7ix\eta+28t(\eta^2+8\eta^4\nu))+a^5(-15+4t\eta(56i\eta+15\delta+336i\eta^3\nu\nonumber\\&&+8\eta^2\delta\nu))-4a^4(10x\eta^2+5i\delta+4t\eta^2(7i\eta+2\delta+72i\eta^3\nu+32\eta^2\delta\nu)),\nonumber\\
	G_3&=&-160ia^8t\eta\nu+8a^7t\eta(-26i\eta+15\delta)\nu+16a^6t\eta(-5i-34i\eta^2\nu+16\eta\delta\nu)+2a^3\eta(156it\eta^3\nonumber\\&&+25i\delta-2\eta^2(13x+4t\delta)+3\eta(1-5ix\delta)+1248it\eta^5\nu-64t\eta^4\delta\nu)-4a^2\eta^2(16it\eta^3-13i\delta\nonumber\\&&+\eta^2(-6x+56t\delta)+8\eta(1+2ix\delta)+128it\eta^5\nu+448t\eta^4\delta\nu)-8\eta^4(i\eta+\delta)(-i+2ix\eta+8t(\eta^2\nonumber\\&&+8\eta^4\nu))-4a\eta^3(i\eta+\delta)(-8i+13ix\eta+52t(\eta^2+8\eta^4\nu))+4a^4(-10x\eta^2+5i\delta+4t\eta^2(13i\eta\nonumber\\&&+8\delta+108i\eta^3\nu+68\eta^2\delta\nu))+a^5(-15+4t\eta(-26i\eta+15\delta-156i\eta^3\nu+172\eta^2\delta\nu)),\nonumber\\
	G_4&=&8a^7t\eta(26i\eta+15\delta)\nu+16a^6t\eta(5i+34i\eta^2\nu+16\eta\delta\nu)+4a^2\eta^2(16it\eta^3+13i\delta-2\eta^2(3x+28t\delta)\nonumber\\&&+8\eta(1-2ix\delta)+128it\eta^5\nu-448t\eta^4\delta\nu)-2a^3\eta(156it\eta^3-25i\delta+\eta^2(-26x+8t\delta)\nonumber\\&&+3\eta(1+5ix\delta)+1248it\eta^5\nu+64t\eta^4\delta\nu)-8\eta^4(-i\eta+\delta)(-i+2ix\eta+8t(\eta^2+8\eta^4\nu))\nonumber\\&&-4a\eta^3(-i\eta+\delta)(-i+2ix\eta+8t(\eta^2+8\eta^4\nu))-4a\eta^3(-i\eta+\delta)(-8i+13ix\eta+52t(\eta^2+8\eta^4\nu))\nonumber\\&&+4a^4(10x\eta^2+5i\delta+4t\eta^2(-13i\eta+8\delta-108i\eta^3\nu+68\eta^2\delta\nu))+a^5(15+4t\eta(26i\eta+15\delta\nonumber\\&&+156i\eta^3\nu+172\eta^2\delta\nu)),\nonumber\\
	B_{bp2}&=&\frac{4a^4(5a+4\eta)^2}{\delta^6}e^{-8t\eta\delta(1+2a^2\nu+4\eta^2\nu)}+\frac{4a^4(5a+4\eta)^2}{\delta^6}e^{8t\eta\delta(1+2a^2\nu+4\eta^2\nu)}+\frac{4i\eta^4}{\delta^7}(-24a^3+8a\eta(3\eta\nonumber\\&&+5i\delta)+a^2(-30\eta+7i\delta)+2\eta^2(15\eta+17i\delta))e^{-4ix\delta}+\frac{4a\eta^4}{\delta^7}(-24ia^2+7a\delta+8\eta(3i\eta+5\delta))e^{4ix\delta}\nonumber\\&&-\frac{8\eta^5}{\delta^7}(-15ia^2+\eta(15i\eta+17\delta))e^{4ix\delta}+\frac{8a\eta}{\delta^5}H_1e^{2\delta(ix+2t\eta(1+2a^2\nu+4\eta^2\nu))}\nonumber\\&&-\frac{8a\eta}{\delta^5}H_2e^{-2\delta(ix+2t\eta(1+2a^2\nu+4\eta^2\nu))}+\frac{8a\eta}{\delta^5}H_3e^{2\delta(-ix+2t\eta(1+2a^2\nu+4\eta^2\nu))}\nonumber\\&&+\frac{8a\eta}{\delta^5}H_4e^{-2\delta(-ix+2t\eta(1+2a^2\nu+4\eta^2\nu))}+\frac{8a}{\delta^6}H_5,\nonumber\\
	H_1&=&160a^6t\eta\nu+8a^5t\eta(41\eta-15i\delta)\nu+16a^4t\eta(5+50\eta^2\nu-6i\eta\delta\nu)+4\eta^2(-40t\eta^3-5\delta\nonumber\\&&+2i\eta^2(5x+12t\delta)+\eta(-3i+6x\delta)-320t\eta^5\nu+192it\eta^4\delta\nu)+2a\eta(-164t\eta^3-25\delta\nonumber\\&&+i\eta^2(41x+60t\delta)+15\eta(-i+x\delta)-1312t\eta^5\nu+480it\eta^4\delta\nu)+a^3(15i+4t\eta(41\eta\nonumber\\&&-15i\delta)(1+8\eta^2\nu))+a^2(40ix\eta^2-20\delta-16it\eta^2(-5i\eta+3\delta)(1+8\eta^2\nu)),\nonumber\\
	H_2&=&160a^6t\eta\nu+8a^5t\eta\nu(41\eta+15i\delta)\nu+16a^4t\eta(5+50\eta^2\nu+6i\eta\delta\nu)-4\eta^2(40t\eta^3-5\delta\nonumber\\&&-2i\eta^2(5x-12t\delta)+\eta(3i+6x\delta)+320t\eta^5\nu+192it\eta^4\delta\nu)-2a\eta(164t\eta^3-25\delta\nonumber\\&&-i\eta^2(41x-60t\delta)+15\eta(i+x\delta)+1312t\eta^5\nu+480it\eta^4\delta\nu)+a^3(15i+4t\eta(41\eta+15i\delta)\nonumber\\&&\times(1+8\eta^2\nu))+4a^2(10ix\eta^2+5\delta+4it\eta^2(5i\eta+3\delta)(1+8\eta^2\nu)),\nonumber\\
\end{eqnarray}
\begin{eqnarray}
	H_3&=&160a^6t\eta\nu+8a^5t\eta(41\eta+15i\delta)\nu+16a^4t\eta(5+50\eta^2\nu+6i\eta\delta\nu)-4\eta^2(40t\eta^3+5\delta\nonumber\\&&+2i\eta^2(5x+12t\delta)-3\eta(i+2x\delta)+320t\eta^5\nu+192it\eta^4\delta\nu)-2a\eta(164t\eta^3+25\delta+i\eta^2(41x\nonumber\\&&+60t\delta)-15\eta(i+x\delta)(1+8\eta^2\nu))+4ia^2(-10x\eta^2+5i\delta+4t\eta^2(5i\eta+3\delta)(1+8\eta^2\nu)),\nonumber\\
	H_4&=&-160a^6t\eta\nu+8ia^5t\eta(41i\eta+15\delta)\nu+6a^4t\eta(5+50\eta^2\nu-6i\eta\delta\nu)+4\eta^2(40t\eta^3-5\delta\nonumber\\&&+2i\eta^2(5x-12t\delta)+\eta(-3i+6x\delta)+320t\eta^5\nu-192it\eta^4\delta\nu)+2a\eta(164t\eta^3-25\delta\nonumber\\&&+i\eta^2(41x-60t\delta)+15\eta(-i+x\delta)+1312t\eta^5\nu-480it\eta^4\delta\nu)+4a^2(10ix\eta^2-5\delta+4t\eta^2(5\eta\nonumber\\&&+3i\delta)(1+8\eta^2\nu))+ia^3(15+4t\eta(41i\eta+15\delta)(1+8\eta^2\nu)),\nonumber\\
	H_5&=&\frac{-128\eta^6}{(-a^2+\eta^2)^3}(1+8x^2\eta^2+128\eta^4(t+8t\eta^2\nu)^2)+3200a^{11}t^2\eta^2\nu^2+5120a^{10}t^2\eta^3\nu^2\nonumber\\&&+5120a^8t^2\eta^3\nu(1+7\eta^2\nu)+128a^9t^2\eta^2\nu(25+191\eta^2\nu)-1280a^6t^2\eta^3(-1-4\eta^2\nu+32\eta^4\nu^2)\nonumber\\&&-32a^7t^2\eta^2(-25-164\eta^2\nu+352\eta^4\nu^2)+32a^2\eta^4(-3x+10x^2\eta+320\eta^3(t+8t\eta^2\nu)^2)\nonumber\\&&+a\eta^4(-43+120x\eta-72x^2\eta^2+896\eta^4(t+8t\eta^2\nu)^2)-40a^4\eta(-1+32t^2\eta^4(5+72\eta^2\nu\nonumber\\&&+256\eta^4\nu^2))+2a^3\eta^2(17-60x\eta+100x^2\eta^2+128t^2\eta^4(15+256\eta^2\nu+1088\eta^4\nu^2))-a^5(-25\nonumber\\&&+32t^2\eta^4(109+1736\eta^2\nu+6912\eta^4\nu^2))-8\eta^5(-12x\eta+40x^2\eta^2+5(1+128\eta^4(t+8t\eta^2\nu)^2)).\nonumber\\
\end{eqnarray}
%%%%%%%%%%%%%%%%%%%%%%
\section*{Acknowledgments}

NVP wishes to thank IISc, Bangalore, for providing a fellowship in the form of  Research Associateship. SM thanks RUSA 2.0 project for providing a fellowship to carry out this work.  The work of MS forms part of a research project sponsored by NBHM, Government of India, under the Grant No. 02011/20/2018 NBHM (R.P)/R\&D II/15064.  The work of GR was supported by Centre for Advanced Study, UGC grant.
%%%%%%%%%%%%%%%%%%%%%%%%%%%
\section*{Authors Contributions}
All the authors contributed equally to the preparation of this manuscript.
\section*{Data Availability Statement}
The data that support the findings of this study are available within the article.

\end{document}